\documentclass[preprint,showpacs,amsmath,amssymb,aps,prd,nofootinbib]{revtex4}
\usepackage{epsfig}
\begin{document}
\draft
\title{\mbox{}\\[10pt]
 $\mu-\tau$ Symmetry  and
 Radiatively Generated Leptogenesis}

\author{Y. H. Ahn$^{1,}$\footnote{E-mail:
        yhahn@cskim.yonsei.ac.kr},
        Sin Kyu Kang$^{2,}$\footnote{E-mail:
        skkang1@sogang.ac.kr},
        C. S. Kim$^{1,}$\footnote{E-mail:
        cskim@yonsei.ac.kr} and
        Jake Lee$^{2,}$\footnote{E-mail:
         jilee@sogang.ac.kr}}

\address{$^{1}$  Department of Physics, Yonsei
        University, Seoul 120-749, Korea\\
        $^{2}$ Center for Quantum Spacetime, Sogang University,
        Seoul 121-742, Korea}



\date{\today}
\begin{abstract}

We consider a $\mu-\tau$ symmetry in neutrino sectors realized at
the GUT scale in the context of a seesaw model. In our scenario, the
exact $\mu-\tau$ symmetry realized in the basis where the charged
lepton and heavy Majorana neutrino mass matrices are diagonal
leads to vanishing lepton asymmetries. We find that, in the
minimal supersymmetric extension of the seesaw model with large
$\tan\beta$, the renormalization group (RG) evolution from the GUT
scale to seesaw scale can induce a successful leptogenesis even
without introducing any symmetry breaking terms by hand, whereas
such RG effects lead to tiny deviations of $\theta_{23}$ and
$\theta_{13}$ from $\pi/4$ and zero, respectively. It is shown
that the right amount of the baryon asymmetry $\eta_B$ can be
achieved via so-called resonant leptogenesis, which can be
realized at rather low seesaw scale with large $\tan\beta$ in our
scenario  so that the well-known gravitino problem is safely
avoided.

\end{abstract}
\pacs{14.60.Pq, 11.30.Fs, 11.10.Hi, 98.80.Cq, 13.35.Hb} \maketitle

\section{Introduction}

Recent precise neutrino experiments appear to show robust
evidence for the neutrino oscillation. The present neutrino
experimental data \cite{atm,SK2002,SNO} exhibit that the
atmospheric neutrino deficit points toward a maximal mixing
between the tau and muon neutrinos. However, the solar neutrino
deficit favors a not-so-maximal mixing between the electron and
muon neutrinos. In addition, although we do not have yet any firm
evidence for the neutrino oscillation arisen from the 1st and 3rd
generation flavor mixing, there is a bound on the mixing element
$U_{e3}$ from CHOOZ reactor experiment, $|U_{e3}|<0.2$
\cite{chooz}. Although neutrinos have gradually revealed their
properties in various experiments since the historic
Super-Kamiokande confirmation of neutrino oscillations \cite{atm},
properties related to the leptonic CP violation are completely
unknown yet. To understand in detail the neutrino mixings observed
in various oscillation experiments is one of the most interesting
issues in particle physics. The large values of $\theta_{sol}$ and
$\theta_{atm}$ may  be telling us about  some underlying new
symmetries of leptons which are not present in the quark sector,
and may provide a clue to understanding the nature of quark-lepton
complementarity beyond the standard model.

Recently, there have been some attempts to explain the maximal
mixing of the atmospheric neutrinos and very tiny value of the 3rd
mixing element $U_{e3}$  by introducing some approximate discrete
symmetries \cite{disc1,disc2} or the mass splitting among the
heavy Majorana neutrinos in the seesaw framework \cite{kangkim}.
In the basis where charged leptons are mass eigenstates, the
$\mu-\tau$ interchange symmetry has become useful in understanding
the maximal atmospheric neutrino mixing and the smallness of
$U_{e3}$ \cite{mu-tau,mu-tau0,mu-tau1,mu-tau2,ahn}. The mass
difference between the muon and the tau leptons, of course, breaks
this symmetry in such a basis. So we expect this symmetry to be an
approximate one, and thus it must hold only for the neutrino
sector. To generate non-vanishing but tiny mixing element
$U_{e3}$, in the literatures \cite{mu-tau2} the authors
introduced $\mu-\tau$ symmetry breaking terms in leptonic mass
matrices by hand at tree level. We have also proposed a scheme for
breaking of $\mu-\tau$ symmetry through an appropriate CP phase in
neutrino Dirac-Yukawa matrix so as to achieve both non-vanishing
$U_{e3}$ and successful leptogenesis \cite{ahn}. In our scheme,
$\mu-\tau$ symmetry breaking factor associated with the CP phase
is essential to achieve both non-vanishing $U_{e3}$ and
leptogenesis. However, besides the soft breaking terms, the
$\mu-\tau$ symmetry is still approximate one in the sense that its
breaking effects in the lepton sector can arise via the
radiative corrections generated by the charged lepton Yukawa
couplings which are not subject to the $\mu-\tau$ symmetry.

In this work we  propose  that the precise $\mu-\tau$
symmetry, imposed in Ref. \cite{ahn}, exists only at high energy
scale such as the GUT scale and  a renormalization group (RG)
evolution from high scale to low scale gives rise to the breaking
of $\mu-\tau$ symmetry in the lepton sector without introducing
any ad hoc soft symmetry breaking terms. However, it  turns out
that the RG effects in the standard model (SM) and even its
minimal supersymmetric extension are quite meager such that the
size of $U_{e3}$ and the deviation of $\theta_{23}$ from the
maximal mixing are tiny\footnote{This is so mainly because our scheme
reflects normal hierarchical light neutrino mass spectrum.}. In
this paper, however, we shall show that such small RG effecs in
supersymmetric seesaw model can lead to successful leptogenesis
which is absent in the exact $\mu-\tau$ symmetry, whereas lepton
asymmetry generated in the context of the SM is too small to
achieve successful leptogenesis. We note that the leptogenesis
realized in our scheme is, in fact, a kind of radiatively induced
leptogenesis which has been discussed in
\cite{radiativeSM,radiativeMSSM}. As will be shown later, in our
scheme both real and imaginary parts of the combination of
neutrino Dirac Yukawa matrix $(Y_\nu Y_\nu^\dag)_{jk}$, which are
needed for leptogenesis, are zero in the limit of the exact
$\mu-\tau$ symmetry at tree level. We note that each of them is
generated via RG effects proportional to $\tan^2\beta$ at low
energy. Thus, the lepton asymmetry generated in our scheme is
proportional to $\tan^4 \beta$ and it can be enhanced by taking
large value of $\tan\beta$. This observation is different from the
results in \cite{radiativeSM,radiativeMSSM}, in which only real
part of $(Y_\nu Y_\nu^\dag)_{jk}$ is radiatively generated and
thus lepton asymmetry is proportional to $\tan^2\beta$.

This paper is organized as follows. In Sec. II, we  present a
supersymmetric seesaw model reflecting $\mu-\tau$ symmetry at a
high energy scale such as the GUT scale. The discussion for RG
evolution from high scale such as the GUT scale to low scale is given
in Sec. III. In Sec. IV, we show how successful leptgenesis can be
radiatively induced in our scheme. Numerical results and
conclusion are given in Sec. V.

\section{Supersymmetric seesaw model with $\mu-\tau$ symmetry realized at the GUT scale}

Let us begin by considering a supersymmetric version of the seesaw
model, which is given as the following leptonic superpotential:
\begin{eqnarray}
W_{\rm lepton}=\widehat{l}_L^c
\textbf{Y}_l\widehat{L}\cdot\widehat{H}_1 +\widehat{N}^c_L
\textbf{Y}_{\nu}\widehat{L}\cdot\widehat{H}_2
-\frac{1}{2}\widehat{N}^{cT}_L\textbf{M}_R\widehat{N}^c_L~,
\label{lagrangian}
\end{eqnarray}
where the family indices have been omitted and $\widehat{L}_j$,
$j=e,\mu, \tau\equiv 1,2,3$ stand for the chiral super-multiplets
of the $SU(2)_L$ doublet lepton fields, $\widehat{H}_{1,2}$ are
the Higgs doublet fields with hypercharge $\mp 1/2$,
$\widehat{N}_{jL}^c$ and $ \widehat{l}^c_{jL}$ are the
super-multiplet of the $SU(2)_L$ singlet neutrino and charged
lepton field, respectively. In the above superpotential,
$\textbf{M}_R$ is the heavy Majorana neutrino mass matrix, and
$\textbf{Y}_l$ and $\textbf{Y}_{\nu}$ are the $3\times 3$ charged
lepton and neutrino Dirac Yukawa matrices, respectively. After
spontaneous symmetry breaking, the seesaw mechanism leads to the
following effective light neutrino mass term,
\begin{eqnarray}
  &&m_{\rm eff}=-\textbf{Y}^{T}_{\nu}\textbf{M}^{-1}_{R}\textbf{Y}_{\nu}\upsilon_2^{2} ~,
  \label{meff}
\end{eqnarray}
where $\upsilon_2$ is the vacuum expectation value of the Higgs
field with positive hypercharge and denoted as
$\upsilon_2=\upsilon\sin\beta$ with $\upsilon\approx174$ GeV.

Let us impose the $\mu-\tau$ symmetry for the neutrino sectors in
the basis in which both the charged lepton mass and heavy
Majorana mass matrices are diagonal, and then the neutrino Dirac-Yukawa
matrix and the heavy Majorana neutrino mass matrix are
given as
\begin{eqnarray} \textbf{Y}_{\nu}={\left(\begin{array}{ccc}
 \textbf{y}_{11} & \textbf{y}_{12} &  \textbf{y}_{12} \\
 \textbf{y}_{12} & \textbf{y}_{22} &  \textbf{y}_{23} \\
 \textbf{y}_{12} & \textbf{y}_{23} &  \textbf{y}_{22} \end{array}\right)},
 ~\textbf{M}_{R}={\left(\begin{array}{ccc}
 M_{1} & 0 &  0 \\
 0 & M_{2} & 0 \\
 0 & 0 & M_{2} \end{array}\right)},
 \label{input1}
 \end{eqnarray}
where the elements $\textbf{y}_{ij}$ of the neutrino Dirac-Yukawa
matrix are all complex in general. As is shown in Ref. \cite{ahn},
the $\mu-\tau$ symmetry imposed as above is responsible for the
neutrino mixing pattern with $\theta_{23}=45^\circ$ and
$\theta_{13}=0^\circ$ after seesawing. Here, we assume that the
above matrices Eq. (\ref{input1}) reflecting the $\mu-\tau$
symmetry are realized at the GUT scale, $Q_{\rm GUT}=2\times 10^{16}$
GeV. As is also shown in \cite{ahn}, the seesaw model based on Eq.
(\ref{input1}) leads to only the normal hierarchical light
neutrino mass spectrum because we take diagonal form of heavy
Majorana neutrino mass matrix \cite{king, ahn}. Thus, the RG
effects on the neutrino mixing matrix $U_{\rm PMNS}$ are expected
to be very small even in the supersymmetric case. However, as will
be shown later, such small RG effects can trigger leptogenesis
which is absent in the case of the exact $\mu-\tau$ symmetry. With
those exact $\mu-\tau$ symmetric structures in the neutrino
sectors, we shall show that a successful leptogenesis could be
achieved solely through the RG running effects between the GUT and
the seesaw scales without being in conflict with experimental low
energy constraints.

\section{Relevant RGE's in MSSM}

 In the minimal supersymmetric standard model (MSSM), the radiative behavior of the heavy Majorana neutrinos
 mass matrix $\textbf{M}_R$
 is dictated by the following RG equation \cite{haba,RG1}:
 \begin{eqnarray}
  \frac{d}{dt}\textbf{M}_R &=& 2[(\textbf{Y}_{\nu}\textbf{Y}^{\dag}_{\nu})\textbf{M}_R
   +\textbf{M}_R(\textbf{Y}_{\nu}\textbf{Y}^{\dag}_{\nu})^{T}],
  \label{RG 0}
 \end{eqnarray}
 where $t=\frac{1}{16\pi^{2}}\ln(Q/Q_{\rm GUT})$ with an arbitrary renormalization scale $Q$.
The RG equation for the neutrino Dirac-Yukawa matrix can be
written as
 \begin{eqnarray}
   \frac{d\textbf{Y}_{\nu}}{dt} &=& \textbf{Y}_{\nu}[(T-3g^{2}_{2}-\frac{3}{5}g^{2}_{1})
   +(Y^{\dag}_{l}Y_{l}+3\textbf{Y}_{\nu}^\dag\textbf{Y}_{\nu})],
  \label{RG 6}
 \end{eqnarray}
where $T={\rm
Tr}(3Y^{\dag}_{u}Y_{u}+\textbf{Y}_{\nu}^\dag\textbf{Y}_{\nu})$,
 and $g_2, g_1$ are the $SU(2)_{L}$ and $U(1)_{Y}$
gauge coupling constants, respectively.

For our convenience, let us re-formulate the RG equation, Eq.
(\ref{RG 0}), in the basis where  $\textbf{M}_R$ is diagonal.
Since $\textbf{M}_R$ is symmetric, it can be diagonalized with a
unitary matrix $V$,
\begin{eqnarray}
  V^{T}\textbf{M}_RV=\text{diag}(M_{1}, M_{2}, M_{3}).
  \label{diagMR}
 \end{eqnarray}
Note that as the structure of mass matrix $\textbf{M}_R$ changes
with the evolution of the scale, that of the unitary matrix $V$
depends on the scale, too. The RG evolution of the unitary matrix
$V(t)$ can be written as
\begin{eqnarray}
  \frac{d}{dt}V &=& VA,
  \label{RGV}
 \end{eqnarray}
 where matrix $A$ is anti-hermitian, $A^\dag = -A$, due to the unitarity of $V$.
 Then, differentiating Eq. (\ref{diagMR}), we obtain
\begin{eqnarray}
  \frac{dM_{i}\delta_{ij}}{dt}=A^{T}_{ij}M_{j}+M_{i}A_{ij}
    +2\{V^{T}[(\textbf{Y}_{\nu}\textbf{Y}^{\dag}_{\nu})\textbf{M}_R
     +\textbf{M}_R(\textbf{Y}_{\nu}\textbf{Y}^{\dag}_{\nu})^{T}]V\}_{ij}.
 \label{RGdiagMR}
 \end{eqnarray}
It immediately follows from the anti-hermiticity of $A$ that
$A_{ii}=0$ in Eq. (\ref{RGdiagMR}).
Absorbing the unitary transformation into the neutrino  Dirac-Yukawa coupling
\begin{eqnarray}
Y_{\nu}\equiv V^{T}\textbf{Y}_{\nu}~, \label{YinDiagMR}
 \end{eqnarray}
 the real diagonal part of Eq. (\ref{RGdiagMR}) becomes
\begin{eqnarray}
  \frac{dM_{i}}{dt}=4M_{i}(Y_{\nu}Y^{\dag}_{\nu})_{ii}~.
  \label{RGMi}
 \end{eqnarray}
On the other hand, the off-diagonal part of Eq. (\ref{RGdiagMR})
leads to
 \begin{eqnarray}
  A_{jk}=2\frac{M_{k}+M_{j}}{M_{k}-M_{j}}{\rm Re}[(Y_{\nu}Y^{\dag}_{\nu})_{jk}]
   +2i\frac{M_{j}-M_{k}}{M_{j}+M_{k}}{\rm Im}[(Y_{\nu}Y^{\dag}_{\nu})_{jk}], ~(j\neq k).
  \label{RGAij}
 \end{eqnarray}
Note that the real part of $A_{jk}$ is singular for the degenerate
cases with $M_j=M_k$, and  the RG equation for $Y_\nu$ in
$\textbf{M}_R$ diagonal basis is written as
 \begin{eqnarray}
   \frac{dY_{\nu}}{dt} &=& Y_{\nu}[(T-3g^{2}_{2}-\frac{3}{5}g^{2}_{1})
        +(Y^{\dag}_{l}Y_{l}+3Y^{\dag}_{\nu}Y_{\nu})]+A^{T}Y_{\nu}.
  \label{RGYinDiagMR}
 \end{eqnarray}
The singularity in ${\rm Re}[A_{jk}]$ can be eliminated with the
help of an appropriate rotation between degenerate heavy Majorana
neutrino states. Such a rotation does not change any physics and
it is equivalent to absorbing the rotation matrix $R$  into the
neutrino Dirac-Yukawa matrix $Y_\nu$,
\begin{eqnarray}
  Y_{\nu}\rightarrow \widetilde{Y}_{\nu}=RY_{\nu},
  \label{transform}
 \end{eqnarray}
where the matrix $R$, particularly rotating 2nd and 3rd generations
of heavy Majorana neutrinos, can be parameterized as
\begin{eqnarray}
R(x)= \left(
\begin{array}{ccc}
 1 & 0 & 0 \\
 0 & \cos x & \sin x \\
 0 & -\sin x & \cos x \\
\end{array}
\right).\label{R1}
\end{eqnarray}
Then, the singularity in the real part of $A_{jk}$ is indeed removed
when the rotation angle $x$ is taken to fulfill the
condition,
\begin{eqnarray}
 {\rm Re}[(\widetilde{Y}_{\nu}\widetilde{Y}^{\dag}_{\nu})_{jk}]=0,
 ~~\text{for any pair}~j,k ~\text{corresponding to}~M_{j}=M_{k}.
\label{singularity2}
\end{eqnarray}

For our purpose, let us parameterize $\textbf{Y}_{\nu}$ at the GUT
scale as follows:
 \begin{eqnarray}
  Y_{\nu}&=& d{\left(\begin{array}{ccc}
 \rho e^{i\varphi_{11}} &  \omega e^{i\varphi_{12}} &  \omega e^{i\varphi_{12}} \\
 \omega e^{i\varphi_{12}} & \kappa  e^{i\varphi_{22}}& e^{i\varphi_{23}} \\
 \omega e^{i\varphi_{12}} & e^{i\varphi_{23}}  & \kappa e^{i\varphi_{22}}
 \end{array}\right)},
 \label{input2}
 \end{eqnarray}
where $\varphi_{ij}$ denote CP phases in $Y_\nu$, and  define the
following useful hermitian parameter
\begin{eqnarray}
  H\equiv(Y_{\nu}Y^{\dag}_{\nu})=d^2
    \left(\begin{array}{ccc}
  H_{11}   &  H_{12}  & H_{12}  \\
  H_{12}^\ast   &  H_{22}  & H_{23} \\
  H_{12}^\ast   &  H_{23}  & H_{22}  \\
\end{array}
\right),
 \label{H}
 \end{eqnarray}
where
\begin{eqnarray}
H_{11} &=& \rho^2 + 2\omega^2,\nonumber\\
H_{12} &=& \rho\omega
e^{i(\varphi_{11}-\varphi_{12})}+\omega\kappa
e^{i(\varphi_{12}-\varphi_{22})}
          +\omega e^{i(\varphi_{12}-\varphi_{23})},\nonumber\\
H_{22} &=& \omega^2 + \kappa^2 +1,\nonumber\\
H_{23} &=& \omega^2 +
2\kappa\cos(\varphi_{22}-\varphi_{23})\label{para}.
\end{eqnarray}
As shown in \cite{ahn},  the hermitian parameter $H$ in the limit
of the exact $\mu-\tau$ symmetry leads to vanishing lepton
asymmetry which is disastrous for successful leptogenesis. To
generate non-vanishing lepton asymmetry, we need to break the
exact degeneracy of the masses of 2nd and 3rd heavy Majorana
neutrinos and the $\mu-\tau$ symmetric texture of $Y_{\nu}$
proposed in Eq. (\ref{input2}). In our scenario, as will be shown
later, only the RG evolution, without including any ad hoc soft
breaking terms, is responsible for such a breaking required for
successful leptogenesis.

For our $\mu-\tau$ symmetric $Y_\nu$ given in Eq. (\ref{input2}),
the angle satisfying the condition Eq. (\ref{singularity2}) is
$x=\pm \pi/4$.  Without a loss of generality, taking $x=\pi/4$, we
obtain
 \begin{eqnarray}
  \widetilde{H}\equiv(\widetilde{Y}_{\nu}\widetilde{Y}^{\dag}_{\nu})= RHR^{T}=
     \left(\begin{array}{ccc}
  H_{11}   &  \sqrt{2}H_{12}  & 0  \\
  \sqrt{2}H_{12}^\ast  &  H_{23}+H_{22}   & 0 \\
  0 &  0  & -H_{23}+H_{22}  \\
\end{array}
\right).
 \label{Htildeplus}
 \end{eqnarray}
It is obvious from Eq. (\ref{Htildeplus}) that ${\rm
Re}[\widetilde{H}_{23(32)}]=0$ and thus the singularity in
$A_{23(32)}$ does not appear. We also note that ${\rm
Im}[\widetilde{H}_{23(32)}]=0$, which is due to the $\mu-\tau$
symmetric structure of $\widetilde{H}$~\footnote{In Ref.
\cite{radiativeSM,radiativeMSSM}, the authors considered the
radiatively induced leptogenesis based on arbitrary textures of
neutrino Dirac-Yukawa matrix for which ${\rm
Im}[\widetilde{H}_{23}]$ needs not to be zero in general.}.
As will be shown later, since the CP asymmetry required for
leptogenesis is proportional to ${\rm
Im}[\widetilde{H}_{23}^2]=2{\rm Re}[\widetilde{H}_{23}] {\rm
Im}[\widetilde{H}_{23}]$, both real and imaginary parts of
$\widetilde{H}_{23}$ should be nonzero for successful
leptogenesis. In this work, we shall show that non-vanishing
values of them can be generated through the RG evolution.

Now, let us consider RG effects which may play an important role in
successful leptogenesis. First, the parameter $\delta_N
=1-M_3/M_2$ reflecting the mass splitting of the degenerate heavy
Majorana neutrinos is governed by the following RGE which can be
derived from Eq. (\ref{RGMi}),
  \begin{eqnarray}
 \frac{d\delta_{N}}{dt}&=& 4(1-\delta_{N})[\widetilde{H}_{22}-\widetilde{H}_{33}]
 \simeq 8{\rm Re}[H_{23}].
 \label{deltaN}
 \end{eqnarray}
The solution of the RGE (\ref{deltaN}) is approximately given by
\begin{eqnarray}
  \delta_{N} &=&
  8d^{2}\{\omega^{2}+2\kappa\cos(\varphi_{22}-\varphi_{23})\} \cdot t,
\label{deltaN1}
 \end{eqnarray}
 where we used Eq. (\ref{para}).
Note that the radiative splitting of degenerate heavy Majorana
neutrinos masses depends particularly on the phase difference,
$\varphi_{22}-\varphi_{23}$.

RGE of the parameter $\widetilde{H}$ is written as
\begin{eqnarray}
  \frac{d\widetilde{H}}{dt} =
  2[(T-3g^{2}_{2}-\frac{3}{5}g^{2}_{1})\widetilde{H}
   +\widetilde{Y}_{\nu}(Y^{\dag}_{l}Y_{l})\widetilde{Y}^{\dag}_{\nu}
   +3\widetilde{H}^2]
   +A^{T}\widetilde{H}+\widetilde{H}A^{\ast}.
  \label{RGHtilde}
 \end{eqnarray}
Considering the structure of $\widetilde{H}$ in Eq.
(\ref{Htildeplus}), up to non-zero leading contributions in the
right side of Eq. (\ref{RGHtilde}), RGE of $\widetilde{H}_{23}$ is
given by
\begin{eqnarray}
  \frac{d{\rm Re}[\widetilde{H}_{23}]}{dt} &\simeq& y_\tau^2{\rm Re}[(\widetilde{Y}_{\nu 23}\widetilde{Y}_{\nu
   33}^\ast)],\nonumber\\
  \frac{d{\rm Im}[\widetilde{H}_{23}]}{dt} &\simeq&
   2{\rm Im}[(\widetilde{Y}_{\nu}Y^{\dag}_{l}Y_{l}\widetilde{Y}^{\dag}_{\nu})_{23}]
   \simeq
   2y_\tau^2{\rm Im}[(\widetilde{Y}_{\nu 23}\widetilde{Y}_{\nu
   33}^\ast)].
  \label{RGHtilde23}
 \end{eqnarray}
In terms of the parameters in Eq. (\ref{para}), the radiatively
generated $\widetilde{H}_{23}$ is given approximately by
\begin{eqnarray}
  {\rm Re}[\widetilde{H}_{23}] &\simeq& y_\tau^2 d^2 \frac{\kappa^2-1}{2} \cdot t , \nonumber\\
  {\rm Im}[\widetilde{H}_{23}] &\simeq& 2y_\tau^2 d^2 \kappa\sin(\varphi_{23}-\varphi_{22}) \cdot t .
\label{Htilde23ReIm}
 \end{eqnarray}
Interestingly enough, the radiatively generated ${\rm
Im}[\widetilde{H}_{23}]$ is proportional to
$\sin(\varphi_{23}-\varphi_{22})$.

\section{Radiatively Induced Resonant Leptogenesis}

When two lightest heavy Majorana neutrinos are nearly degenerate,
the CP asymmetry through their decays gets dominant contributions
from self-energy diagrams and can be written as
\cite{lepto,lepto2,lepto_self,pilaftsis}
 \begin{eqnarray}
  \varepsilon_{i}=\frac{\Gamma(N_{i}\rightarrow l\varphi)
  -\Gamma(N_{i}\rightarrow \overline{l}\varphi^{\dag})}{\Gamma(N_{i}\rightarrow l\varphi)
  +\Gamma(N_{i}\rightarrow\overline{l}\varphi^{\dag})}
  \simeq
   \sum_{k\neq
  i}\frac{{\rm Im}[(Y_{\nu}Y^{\dag}_{\nu})^{2}_{ik}]}{16\pi(Y_{\nu}Y^{\dag}_{\nu})_{ii}\delta_{N,ik}}
  \Big(1+\frac{\Gamma^{2}_{k}}{4M^{2}_{i}\delta^{2}_{N,ik}}\Big)^{-1},
 \label{epsilon}
 \end{eqnarray}
where $\Gamma_{k}$ is the tree-level decay width of the $k$-th
right-handed neutrino,
\begin{eqnarray}
  \Gamma_{k}= \frac{(Y_{\nu}Y^{\dag}_{\nu})_{kk}M_{k}}{8\pi},
 \end{eqnarray}
and $\delta_{N,ik}$ is a parameter which denotes the degree of the
mass splitting between two degenerate heavy Majorana neutrinos,
\begin{eqnarray}
  \delta_{N,ik}\equiv 1-\frac{M_{k}}{M_{i}}.
  \label{degeneracy1}
 \end{eqnarray}
 As shown in \cite{ahn}, the neutrino Dirac-Yukawa matrix and the heavy Majorana neutrino mass
 matrix given in the forms of Eq. (\ref{input1}) are consistent with neutrino oscillation data only when
$M_1 \gg M_2 \simeq M_3$.
Here we note that  it is rather difficult to realize naturally such
an inverted hierarchy of the heavy  Majorana neutrino mass spectrum in GUT models.
For the mass hierarchy  $M_1 \gg M_2 \simeq M_3$,
the decay of $N_1$ takes place in thermal equilibrium and thus
 the lepton asymmetry required for successful leptogenesis will be accomplished by
 $\varepsilon_{2(3)}$ given as follows:
\begin{eqnarray}
  \varepsilon_{2(3)}=
   \frac{{\rm Im}[(Y_{\nu}Y^{\dag}_{\nu})^{2}_{23}]}{16\pi(Y_{\nu}Y^{\dag}_{\nu})_{22(33)}\delta_{N}}
  \Big(1+\frac{\Gamma^{2}_{3(2)}}{4M^{2}_{2(3)}\delta^{2}_{N}}\Big)^{-1},
 \label{epsilon23}
 \end{eqnarray}
where $\delta_N = \delta_{N,23}$. {}From Eqs.
(\ref{Htildeplus},\ref{deltaN1},\ref{Htilde23ReIm}), the lepton
asymmetry is given by
\begin{eqnarray}
  \varepsilon_{2(3)} &\simeq&  \frac{{\rm Im}[(\widetilde{H}_{23})^2]}{16\pi\widetilde{H}_{22(33)}\delta_N}\nonumber\\
   &\simeq& \frac{y^{4}_{\tau}\kappa(\kappa^{2}-1)\sin(\Delta\varphi)\cdot t}
                 {64\pi \{\omega^{2}+2\kappa\cos(\Delta\varphi)\}\cdot h_{2(3)}},
  \label{epsilon23approx}
 \end{eqnarray}
where $\Delta\varphi \equiv \varphi_{23}-\varphi_{22}$, and two
parameters $h_{2(3)}$ are defined as
\begin{eqnarray}
  h_{2} &=&\widetilde{H}_{22}/d^2= 1+\kappa^{2}+2\omega^{2}+2\kappa\cos(\Delta\varphi),\nonumber \\
  h_{3} &=& \widetilde{H}_{33}/d^2= 1+\kappa^{2}-2\kappa\cos(\Delta\varphi).
  \end{eqnarray}
In Eq. (\ref{epsilon23approx}) we neglected the term containing
the decay width since it  turns out to be very small in our
scenario. Note that due to the opposite sign of the term
$2\kappa\cos(\Delta\varphi)$ in $h_2$ and $h_3$, either $h_{2}$ or
$h_3$ becomes larger depending on the sign of
$\cos(\Delta\varphi)$. This implies that either of the two
degenerate heavy Majorana neutrinos, $N_2$ or $N_3$, would
dominantly contribute to the leptogenesis over two distinct
regions of $\Delta\varphi$. More specifically, for $\Delta\varphi
< 90^\circ$ or $\Delta\varphi > 270^\circ$, $\epsilon_2$ is
dominant over $\epsilon_3$  because of $h_2 \ll h_3$. Otherwise,
$\epsilon_3$ is dominant over $\epsilon_2$. However, in our
scenario as shown in Fig. 1,  only the former case ($\epsilon_2 \gg
\epsilon_3)$ is allowed, mainly because of the experimental
constraint $\Delta m_{sol}^2/m_{atm}^2 \ll 1$, as will be shown
later in detail.

We remark that the radiatively induced lepton asymmetry
$\epsilon_i$ is proportional to $y_\tau^4 =
y_{\tau,SM}^4(1+\tan^2\beta)^2$, and thus for large $\tan\beta$ it
can be highly enhanced and proportional to $\tan^4\beta$. Furthermore,
it has an explicit dependence of the evolution scale $t$. These
two points are different from what was obtained in \cite{radiativeSM,
radiativeMSSM}, where the neutrino Dirac -Yukawa matrix has been
arbitrary chosen so that ${\rm Im}[\widetilde{H}_{23}]$ could be
initially non-zero and the lepton asymmetry became proportional
to $y_\tau^2$ at leading order and, at the same time, the scale
dependence was cancelled out.

The resulting baryon-to-photon ratio is estimated in the context
of MSSM to be
 \begin{eqnarray}
  \eta_{B}\simeq -1.67\times 10^{-2}\sum_{i}\varepsilon_{i}\cdot\kappa_{i},
 \end{eqnarray}
where the efficiency factor $\kappa_{i}$ describes the washout of
the produced lepton asymmetry $\varepsilon_{i}$. The efficiency in
generating the resultant baryon asymmetry is usually controlled by
the parameter defined as
 \begin{eqnarray}
 K_i \equiv \frac{\Gamma_i}{H} = \frac{\tilde{m}_i}{m_\ast},
 \end{eqnarray}
where $H$ is the Hubble constant, and $\widetilde{m}_i$, the
so-called effective neutrino mass, is given by
\begin{eqnarray}
\widetilde{m}_{i}=\frac{[m_{D}m^{\dag}_{D}]_{ii}}{M_{i}},
 \end{eqnarray}
and ${m_\ast}$ is defined as
\begin{eqnarray}
  m_{\ast} &=& \frac{16
  \pi^{\frac{5}{2}}}{3\sqrt{5}}g^{\frac{1}{2}}_{\ast}\frac{\upsilon^{2}}{M_{\rm Planck}}\simeq
  1.08\times10^{-3}\;{\rm eV},
 \end{eqnarray}
 where we adopted $M_{\rm Planck}=1.22\times 10^{19}\;{\rm GeV}$ and
 the effective number of degrees of freedom
$g_{\ast} \simeq g_{\ast{\rm SM}}=106.75$, $g_{\ast{\rm MSSM}}
\simeq 2g_{\ast{\rm SM}}$. Although most analyses on baryogenesis
via leptogenesis conservatively consider $K_i < 1$, much larger
values of $K_i$, even larger than $10^3$, can be tolerated
\cite{pilaftsis}.

{}From the actual numerical calculations, we find that our
scenario resides in the so-called {\it strong washout regime} with
\begin{eqnarray}
K_2 \gtrsim 1, \;\;\;\; K_3 \gtrsim 10.
 \end{eqnarray}
Thus, for our numerical calculations, we will adopt approximate
expressions of the efficiency factor given for large $K_i$ by
\cite{nielsen},
 \begin{eqnarray}
  \kappa_{i} &\approx& \frac{1}{2\sqrt{K^{2}_{i}+9}}~~~~~~~~~~~~~
                       \text{for}~0\lesssim K_{i}\lesssim 10,\\
  \kappa_{i} &\approx& \frac{0.3}{K_{i}({\rm ln}K_{i})^{0.6}}~~~~~~~~~~~~
                       \text{for}~10\lesssim K_{i}\lesssim
  10^{6}.\nonumber
  \label{washout 2}
 \end{eqnarray}


\section{Numerical Analysis and Discussions}

As can be seen in the approximate expression in Eq.
(\ref{epsilon23approx}), the lepton asymmetry $\epsilon_i$ depends
dominantly on one phase difference,
$\Delta\varphi=\varphi_{23}-\varphi_{22}$, among the phases
assigned in the neutrino Dirac-Yukawa matrix given in the form of
Eq. (\ref{input2}). Therefore, we focus on the phase difference
$\Delta\varphi$ and study how the prediction of $\eta_B$ varies
with the choice of the input values of  $\Delta\varphi$ at the GUT
scale. In order to estimate the RG evolutions of neutrino Dirac-Yukawa
matrix and heavy Majorana neutrino masses from the GUT scale to
the seesaw scale, we numerically solve all the relevant RGE's
presented in \cite{antusch}.


\begin{figure}[b]
\hspace*{-2cm}
\begin{minipage}[t]{6.0cm}
\epsfig{figure=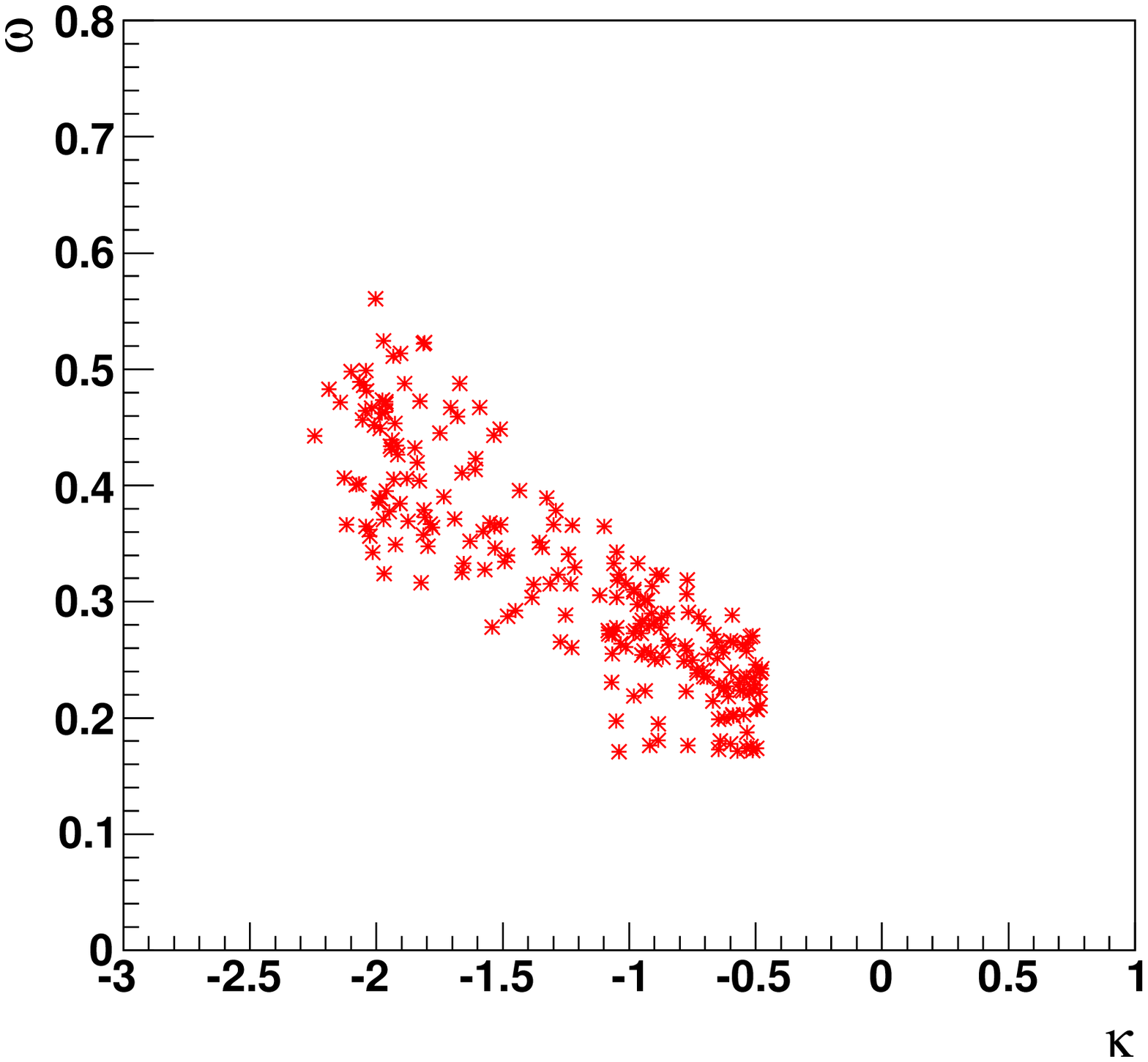,width=8.5cm,angle=0}
\end{minipage}
\hspace*{2.0cm}
\begin{minipage}[t]{6.0cm}
\epsfig{figure=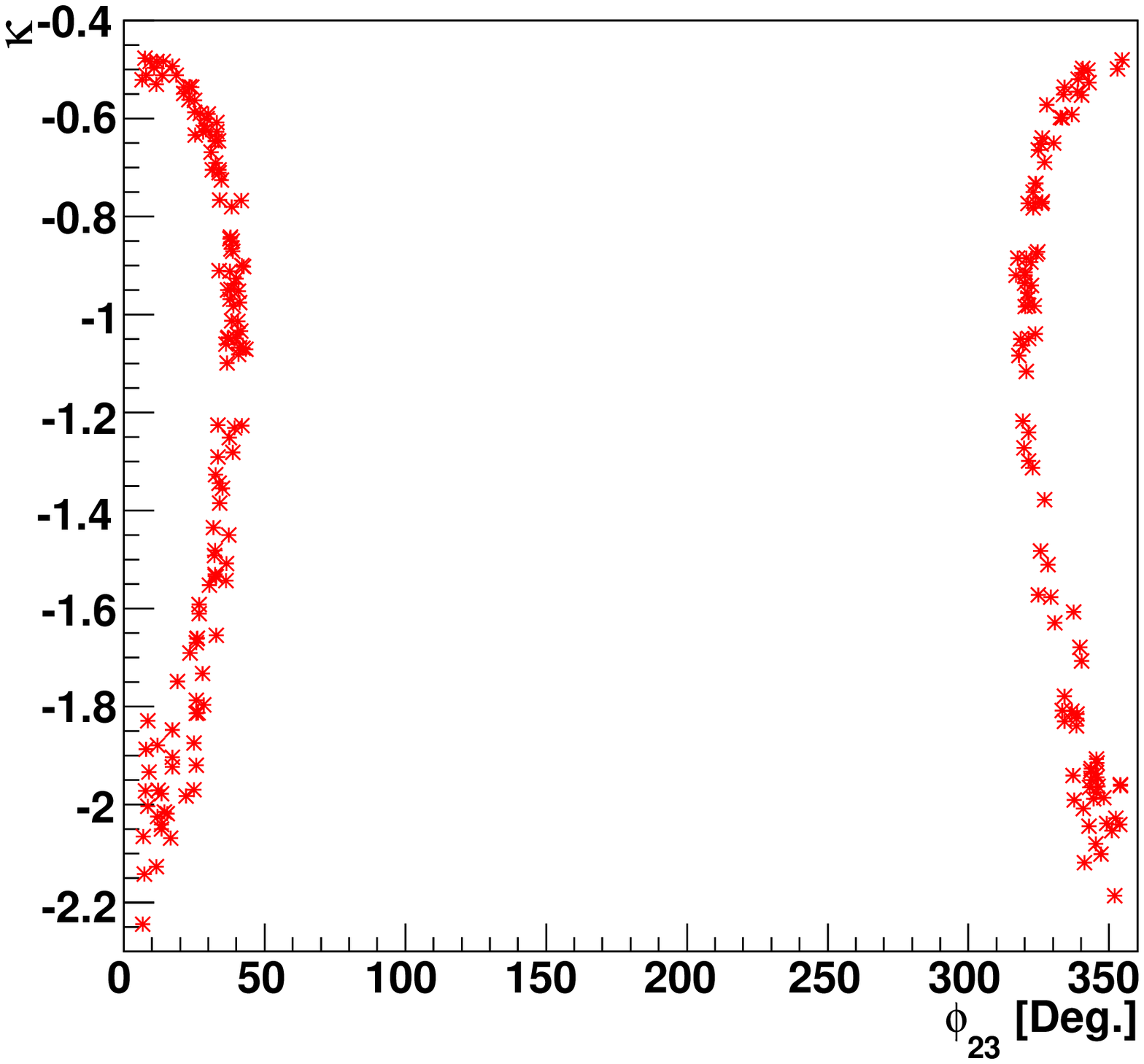,width=8.5cm,angle=0}
\end{minipage}
\caption{\label{Fig1}Parameter regions allowed by the $3\sigma$
experimental constraints in Eq.~(\ref{exp_bound}) for
$M_1=10^{13}\;{\rm GeV}$, $M_2=10^{6}\;{\rm GeV}$ and
$\tan\beta=25$.}
\end{figure}

In our numerical calculation of the RG running effects, we first
fix the values of two masses of heavy Majorana neutrinos with
hierarchy $M_1 \gg M_2$ and $\tan\beta$, then we solve the RGE's
by varying input values of all the parameter space $\{
d,\kappa,\omega,\rho,\Delta\varphi\}$ given at the GUT scale. Then
finally we determine the parameter space allowed by low energy
neutrino experimental data. At present, we have five experimental
data, which are taken as low energy constraints in our numerical
analysis, given at $3\sigma$  by \cite{maltoni},
\begin{eqnarray}
&&29.3^\circ < \theta_{12} < 39.2^\circ,\quad 35.7^\circ <
\theta_{23} < 55.6^\circ,\quad
0^\circ < \theta_{13} < 11.5^\circ,\nonumber\\
&&7.1<\Delta m^2_{21}[10^{-5}{\rm eV}^2]<8.9,\quad 2.0<\Delta
m^2_{31}[10^{-3}{\rm eV}^2]<3.2\label{exp_bound}.
\end{eqnarray}

Using the results of the RG evolutions, we estimate the lepton
asymmetry for the allowed parameter space from these low energy
experimental constraints. In Fig.~\ref{Fig1}, we show the
parameter regions constrained by the experimental data given in
Eq. (\ref{exp_bound}). The two figures exhibit how the parameters
$\kappa$ and $\omega$ are correlated and how $\kappa$ depends on
the phase difference $\Delta\varphi$, respectively. Here we
adopted $M_1=10^{13}\;{\rm GeV}$, $M_2=10^{6}\;{\rm GeV}$ and
$\tan\beta=25$ as inputs, so that the gravitino could not be
overproduced in early Universe\footnote{We note that the mass of
$M_2$ can be as light as $10^3$ GeV in our scenario.}.

We find that due to the mass hierarchy of the heavy Majorana
neutrinos $M_1 \gg M_2$, the RG running effects depend very weakly
on the parameter $\rho$ in our analysis.


\begin{figure}[b]
\hspace*{-2cm}
\begin{minipage}[t]{6.0cm}
\epsfig{figure=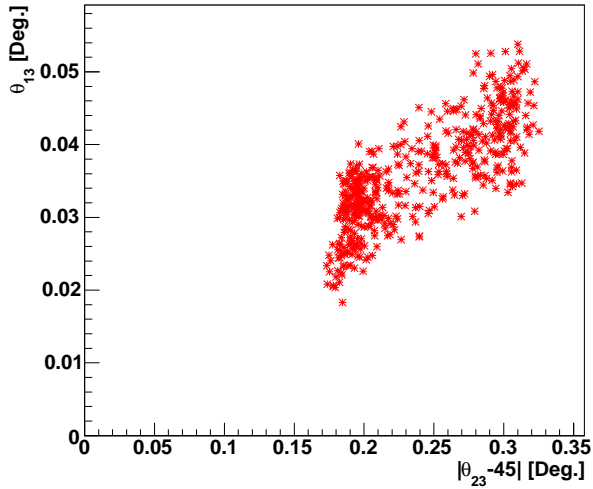,width=8.5cm,angle=0}
\end{minipage}
\caption{\label{Fig2} Radiatively generated deviations of
$\theta_{13}$ and $\theta_{23}$ from the $\mu-\tau$ symmetric
prediction, ($\theta_{13}=0$ and $\theta_{23}=45^\circ$), for the
same parameter space used in Fig.1.}
\end{figure}

As mentioned earlier, since our scenario allows only the normal
hierarchical spectrum of light neutrino masses, there are very
tiny deviations of the mixing angles arising from the RG
evolutions,  even in the supersymmetric case with large
$\tan\beta$. In Fig.~\ref{Fig2}, we show the deviations of
$\theta_{13}$ and $\theta_{23}$ from their $\mu-\tau$ symmetric
initial values at the GUT scale, {\it i.e.} $\theta_{13}=0$ and
$\theta_{23}=45^\circ$, resulting from  the RG evolution. It turned
out from our numerical estimate that the radiatively generated
deviations of $\theta_{13}$ and $\theta_{23}$ from the initial
angles are at most $0.2^\circ$ and $1.2^\circ$, respectively, even
for $\tan\beta \sim 50$. In addition, there can exist radiative
corrections associated with low energy supersymmetric threshold
effects, which might be important in some cases \cite{threshold}.
The typical size of the flavor diagonal threshold corrections
denoted by  $I^{TH}_{\alpha} \sim g_2^2/(32\pi^2) f_{\alpha}$
($\alpha=e,\mu,\tau$) with a loop function $f_{\alpha}$
\cite{threshold} is of order $10^{-3}$ and corresponding
additional deviations of mixing angles are at most less than
$0.1^\circ$. However, in the case that either $I^{TH}_{\tau}$ or
$I^{TH}_{\mu}$ is dominant and its size reaches  maximally
allowed value 0.03 \cite{chan}, the additional deviations of the
mixing angles can be $\delta \theta_{13}\sim 0.2^\circ$ and
$\delta \theta_{23}\sim 1^\circ$.


\begin{figure}[tb]
\hspace*{-2cm}
\begin{minipage}[t]{6.0cm}
\epsfig{figure=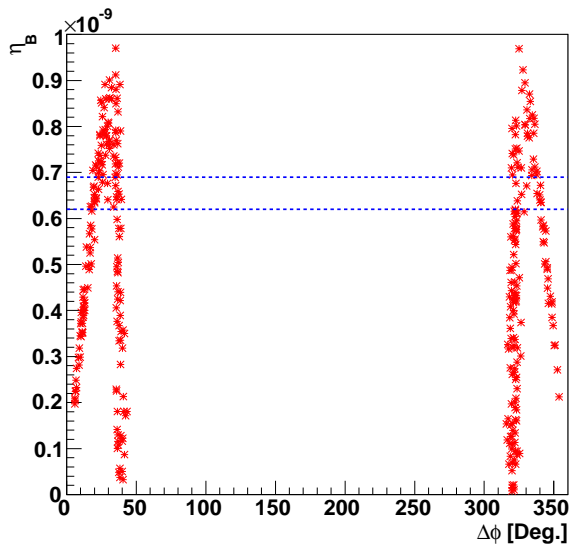,width=8.5cm,angle=0}
\end{minipage}
\caption{\label{Fig3}Predictions for the baryon asymmetry $\eta_B$
for the same parameter space as in Fig. 1. The horizontal lines are
the current bounds from the CMB observations.}
\end{figure}

In Fig. 3, we present the predictions for $\eta_B$ as a function
of the phase difference $\Delta\varphi$ imposed initially at the GUT
scale. The horizontal lines correspond to the current bounds from
the CMB observations \cite{cmb}:
\begin{eqnarray}
\eta_B^{CMB}=(6.5^{+0.4}_{-0.3})\times 10^{-10} \;\;(1\sigma).
\label{cmb}
\end{eqnarray}


\begin{figure}[tb]
\hspace*{-2cm}
\begin{minipage}[t]{6.0cm}
\epsfig{figure=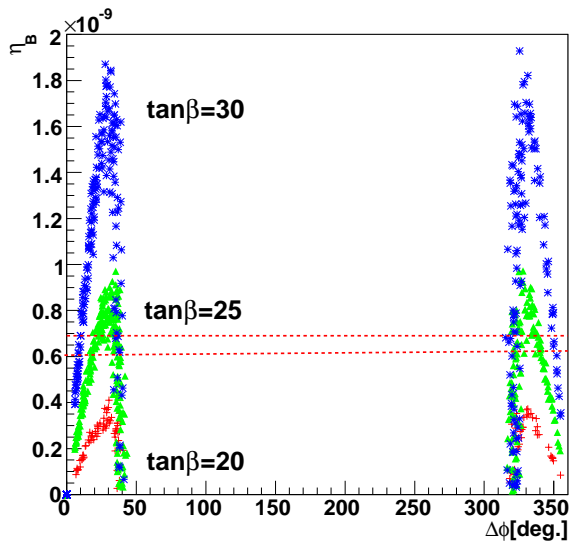,width=8.5cm,angle=0}
\end{minipage}
\caption{\label{Fig4}Predictions for the baryon asymmetry $\eta_B$
for the same parameter space as in Fig. 1. The different colors
stand for the cases with $\tan\beta=20$ (red cross), 25 (green
triangle), 30 (blue star). The horizontal lines are the current
bounds from the CMB observations.}
\end{figure}

In Fig. 4, we show how $\tan\beta$ dependence of the baryon
asymmetry $\eta_B$ varies with the phase difference
$\Delta\varphi$. The different colored points stand for the
results for $\tan\beta=20$ (red cross), 25 (green triangle), 30
(blue star). We see that the predictions of $\eta_B$ get smaller
as $\tan\beta$ decreases. Thus, we can extract lower limit of
$\tan\beta$ from the current observation for $\eta_B$ given in Eq.
(\ref{cmb}), which is parameterized as follows;
\begin{eqnarray}
\tan^4\beta \gtrsim 2\times 10^4 \left[ \frac{(\omega^2+2\kappa
\cos\Delta\varphi)h_2}{\kappa(\kappa^2-1)\sin\Delta\varphi\cdot t}
\right]. \label{tanb4bound}
\end{eqnarray}
Numerically, the maximum peaks correspond to $\Delta\varphi \simeq
30^\circ ~\mbox{and}~~ 330^\circ$ as can be seen in Fig. 4, and
the successful leptogenesis can be achieved in our scenario only
for the value of $\tan\beta$ satisfying
\begin{eqnarray}
\tan\beta \gtrsim 23. \label{tanbbound}
\end{eqnarray}

As a summary, we have considered an exact $\mu-\tau$ symmetry in
neutrino sectors realized at the GUT scale in the context of a seesaw
model. The exact $\mu-\tau$ symmetry, which is realized in the
basis where the charged lepton and heavy Majorana neutrino mass
matrices are diagonal, leads to vanishing lepton asymmetries. We
have shown that, in the minimal supersymmetric extensions of the
seesaw model with large $\tan\beta$, the RG evolution from the GUT
scale to the seesaw scale can induce a successful leptogenesis without
introducing any symmetry breaking terms by hand, whereas such
small RG effects lead to tiny deviations of $\theta_{23}$ and
$\theta_{13}$ from their initial values at the GUT scale, {\it i.e.}
$\theta_{23}=\pi/4$ and $\theta_{13}=0$, respectively. The right
amount of the baryon asymmetry $\eta_B$ has been achieved via
so-called resonant leptogenesis. In our scenario the seesaw scale
can be lowered down to as much as $10^3$ GeV for $\tan\beta=25$
and so the well-known gravitino problem is safely avoided.
\\

\acknowledgments{ \noindent The work of C.S.K. was supported in
part by  CHEP-SRC Program and in part by the Korea Research
Foundation Grant funded by the Korean Government (MOEHRD) No.
KRF-2005-070-C00030. SKK and JL were supported by the SRC program
of KOSEF through CQUeST with Grant No. R11-2005-021. SKK was
supported by the Korea Research Foundation Grant funded by the
Korean Government(MOEHRD) (KRF-2006-003-C00069). YHA was supported
by Brain Korea 21 Program.}

\end{document}